\documentclass[twocolumn]{aa}
\usepackage{graphicx}
\usepackage{natbib}
\usepackage{aalongtable}

\def\ibvs{IBVS}
\def\an{Astronomische Nachrichten}
\def\IAUColloq{IAU Colloq.}
\def\AcA{Acta Astron.}
\def\PublisherKluwer{Dordrecht: Kluwer Academic Publisher}
\def\PublisherASP{San Francisco: Astronomical Society of the Pacific}
\def\PublisherReidel{Dordrecht: D. Reidel Publishing Co.}
\usepackage{txfonts}
\begin{document}
   \title{TV Corvi Revisited: Precursor and Superhump Period Derivative Linked
to the Disk Instability Model}

   \author{M. Uemura\inst{1,2} \and R.E. Mennickent\inst{1} \and 
   R. Ishioka\inst{3} \and A. Imada\inst{3} \and T. Kato\inst{3} \and
   D. Nogami\inst{4} \and R. Stubbings\inst{5} \and S. Kiyota\inst{6}
   \and \\ P. Nelson\inst{7} \and K. Tanabe\inst{8} \and
   B. Heathcote\inst{9} \and G. Bolt\inst{10}} 

   \offprints{M. Uemura}

   \institute{Departamento de F\'{\i}sica, Facultad de Ciencias
              F\'{\i}sicas y Matem\'aticas, Universidad de Concepci\'on,
              Casilla 160-C, Concepci\'on, Chile\\ 
              \email{muemura@cepheid.cfm.udec.cl}
         \and Yukawa Institute for Theoretical Physics, Kyoto
              University, Kyoto 606-8502, Japan
	 \and Department of Astronomy, Kyoto University, Sakyou-ku,
              Kyoto 606-8502, Japan
         \and Hida Observatory, Kyoto University, Kamitakara, Gifu
              506-1314, Japan
         \and 19 Greenland Drive, Drouin 3818, Victoria, Australia
         \and Variable Star Observers League in Japan (VSOLJ), 1-401-810
              Azuma, Tsukuba 305-0031, Japan
         \and RMB 2493, Ellinbank 3820, Australia
         \and Department of Biosphere-Geosphere Systems, Faculty of
              Informatics, Okayama University of Science, Ridaicho 1-1,
              Okayama 700-0005, Japan
         \and Barfold Observatory, Mia Mia, Victoria, Australia
         \and 295 Camberwarra Drive, Craigie 6025, Western Autralia}

   \date{Received, 14 September 2004; accepted, 1 November 2004}

   \abstract{
 We report optical photometric observations of four superoutbursts of
 the short-period dwarf nova TV~Crv.  This object experiences two types
 of superoutbursts; one with a precursor and the other without.  The
 superhump period and period excess of TV~Crv are accurately
 determined to be $0.065028\pm 0.000008$~d and $0.0342\pm 0.0021$,
 respectively.  This large excess implies a relatively large mass
 ratio of the binary components ($M_2/M_1$), though it has a short orbital
 period.  The two types of superoutbursts can be explained by the
 thermal-tidal instability model for systems having large mass ratios.
 Our observations reveal that superhump period derivatives are variable
 in distinct superoutbursts.  The variation is apparently related to the
 presence or absence of a precursor.  We propose that the superhump
 period derivative depends on the maximum disk radius during outbursts.
 We investigate the relationship of the type of superoutbursts and the
 superhump period derivative for known sources.  In the case of
 superoutbursts without a precursor, superhump period derivatives tend
 to be larger than those in precursor-main type superoutbursts, which is
 consistent with our scenario.

   }

   \maketitle
%

\section{Introduction}

SU~UMa-type stars form a sub-group of dwarf novae characterized by the
appearance of long and bright ``superoutbursts'', during which periodic
modulations, ``superhumps'', are observed \citep{war85suuma}.  Superhumps
have periods slightly longer than orbital periods, which can 
be explained by a beat phenomenon of a precessing tidally-distorted
eccentric disk.  According to the tidal instability theory, an accretion
disk becomes unstable against a tidal perturbation from a secondary star
when the disk reaches the 3:1 resonance radius \citep{whi88tidal}.  In
conjunction with the thermal instability model for (normal) dwarf
nova outbursts, the model for superoutbursts is called the
thermal-tidal instability (TTI) model \citep{osa89suuma}.

Superoutbursts are sometimes associated with a precursor typically
lasting one or two days.  This precursor phenomenon is actually
expected from the TTI model.  The precursor is considered to
be a normal outburst leading to an expansion of the accretion disk
over the 3:1 resonance radius and triggering a superoutburst.  Growing
superhumps have been detected during a decay phase from the precursor in
T~Leo \citep{kat97tleo}, V436~Cen \citep{sem80v436cen}, and GO~Com
\citep{ima04gocom}.  These growing superhumps provide evidence for
the TTI model since a system is predicted to reach a supermaximum with
the growth of an eccentric disk.  On the other hand, the original TTI
model cannot explain gradually growing superhumps even after supermaxima
without a precursor, which are also frequently observed 
\citep{sma96superoutburst}.  

\citet{osa03DNoutburst} propose a refinement of the original TTI model with
the idea that the accretion disk can pass the 3:1 resonance radius and
reach the tidal truncation radius.  The dammed matter at the tidal
truncation radius causes a gradual decay without a precursor.  This
refined TTI model predicts that SU~UMa stars having a large mass ratio
($q=M_2/M_1$, where $M_1$ and $M_2$ are the masses of a white dwarf and a
secondary star, respectively) can show both types of superoutbursts,
that is, those with and without a precursor.  This idea should be
examined by observations of the early evolution of superoutbursts and 
superhumps.  

The superhump period ($P_{\rm SH}$) in SU~UMa stars generally decreases
through a superoutburst with a period derivative of order 
$\dot{P}_{\rm SH}/P_{\rm SH}\sim -10^{-5}$
\citep{war85suuma,pat93vyaqr}.  A simple dynamical treatment for the
tidal 
instability shows that the precession rate of the eccentricity wave is
proportional to $r^{1.5}$, where $r$ is the disk radius
\citep{osa85SHexcess}.  The shortening of $P_{\rm SH}$ can,
hence, be understood with the shrink of the disk during a superoutburst.  
Hydrodynamical simulations also show that the precessing eccentricity
wave propagates inward, which causes the period shortening of
superhumps \citep{lub92SH,whi94SH}.

On the other hand, several short-period SU~UMa stars showing positive
$\dot{P}_{\rm SH}/P_{\rm SH}$ have been discovered since mid-90's 
\citep{how96alcom,kat03v877arakktelpucma}.  WZ~Sge-type stars,
in particular, tend to show positive $\dot{P}_{\rm SH}/P_{\rm
SH}$  \citep[e.g.][]{how96alcom,kat97egcnc}.  The situation
becomes more complicated because ultra-short period systems, V485~Cen and 
EI~Psc also show positive $\dot{P}_{\rm SH}/P_{\rm SH}$
\citep{ole97v485cen,uem02j2329}.  These two sources have
quite large mass ratios ($q\sim 0.2$), though WZ~Sge stars have quite
small mass ratio ($q\sim 0.01$).  Based on the discussions for
ordinary negative $\dot{P}_{\rm SH}/P_{\rm SH}$, the positive
$\dot{P}_{\rm SH}/P_{\rm SH}$ has been proposed to arise due to an expansion
of the disk or an outward-propagation of the eccentricity wave
\citep{bab00v1028cyg,kat04egcnc}.  It is, however, poorly
understood why the outward propagation can occur only in the
short-period systems regardless of their mass ratio \citep{ish03hvvir}. 

TV~Crv is known as an SU UMa-type dwarf nova having a short orbital
period of $0.06288\pm 0.00013$~d \citep{wou03tvcrv}.  The historical 
discovery of this object is summarized in \citet{lev90tvcrv}.
\citet{how96tvcrv} reported superhumps with a period of $0.0650\pm
0.0008\;{\rm d}$ from observations of a superoutburst in 1994 June.
This $P_{\rm SH}$ provides a superhump period excess
$\varepsilon=(P_{\rm SH}-P_{\rm orb})/P_{\rm orb}=0.033\pm 0.009$.
This value of the excess implies that TV~Crv may be a peculiar object
regarding its possibly large period excess compared with other short
period systems \citep{pat01SH}.  The error of $\varepsilon$ is,
however, so large that the large $\varepsilon$ is not conclusive. 

Here we report observations of four superoutbursts of TV~Crv.  Our
observations on TV~Crv provide  new clues to understand the superhump
period evolution related to the precursor phenomenon and the TTI
model.  In the next section, we mention our observation systems.  In
Sect.~3, we report detailed behaviour of superoutbursts and
superhumps of TV~Crv.  We then discuss the implication of our results
linked to the TTI model in Sect.~4 and 5.  In Sect.~6, we compare and
discuss our results with those for other known systems.  Finally, we
summarize our findings in Sect.~7. 

\section{Observations}

\begin{table}
  \caption{Journal of observations.}\label{tab:log}
\centering
\begin{tabular}{cccrr}
  \hline\hline
  ID & $T_{\rm start}$ & $\delta T$ & $N$ & Site\\
  \hline
01-01 & 1957.1607 & 4.75 & 361 & Kyoto\\
01-02 & 1958.1373 & 5.28 & 243 & Tsukuba\\
01-03 & 1958.2718 & 2.12 & 176 & Kyoto\\
01-04 & 1959.0578 & 7.21 & 542 & Kyoto\\
01-05 & 1960.3133 & 0.75 &  64 & Kyoto\\
01-06 & 1961.1367 & 4.90 & 374 & Kyoto\\
01-07 & 1961.1470 & 3.04 &  70 & Tsukuba\\
01-08 & 1963.2215 & 1.96 & 138 & Kyoto\\
01-09 & 1964.0792 & 2.78 &  73 & Tsukuba\\
01-10 & 1965.1255 & 5.17 & 436 & Kyoto\\
01-11 & 1965.1483 & 3.81 &  94 & Tsukuba\\
01-12 & 1966.1710 & 0.56 &  39 & Kyoto\\
01-13 & 1969.1343 & 4.44 & 303 & Kyoto\\
02-01 & 2427.9767 & 2.18 & 132 & Kyoto\\
02-02 & 2428.0206 & 1.47 & 119 & Kyoto\\
02-03 & 2428.9583 & 2.23 & 195 & Kyoto\\
02-04 & 2428.9960 & 1.86 &  51 & Kyoto\\
02-05 & 2429.0071 & 1.56 & 147 & Okayama\\
02-06 & 2430.9581 & 1.16 & 103 & Kyoto\\
02-07 & 2430.9627 & 2.34 & 182 & Kyoto\\
02-08 & 2434.9773 & 2.03 & 161 & Kyoto\\
03-01 & 2769.9589 & 6.18 & 327 & Craigie\\
03-02 & 2776.9232 & 1.34 &  71 & Ellinbank\\
03-03 & 2777.0997 & 2.88 &  31 & Kyoto\\
03-04 & 2777.8617 & 2.18 & 111 & Ellinbank\\
03-05 & 2780.1247 & 1.35 &  37 & Kyoto\\
03-06 & 2781.0285 & 1.82 &  86 & Hida\\
04-01 & 3160.9673 & 2.46 & 237 & Kyoto\\
04-02 & 3161.9689 & 2.05 & 142 & Kyoto\\
04-03 & 3162.4693 & 5.84 & 210 & Concepci\'on\\
04-04 & 3163.4698 & 6.18 & 238 & Concepci\'on\\
04-05 & 3164.9970 & 0.70 &  34 & Barfold\\
04-06 & 3166.5550 & 3.81 & 182 & Concepci\'on\\
04-07 & 3167.5709 & 3.40 & 195 & Concepci\'on\\
04-08 & 3168.5882 & 2.86 & 176 & Concepci\'on\\
04-09 & 3169.5472 & 1.67 &  57 & Concepci\'on\\
04-10 & 3169.9654 & 2.63 & 123 & Kyoto\\
04-11 & 3170.9610 & 3.06 & 189 & Kyoto\\
  \hline
  \multicolumn{5}{l}{$T_{\rm start}=$HJD$-$2450000.}\\
  \multicolumn{5}{l}{$\delta T=$Period of observations in hours.}\\
  \multicolumn{5}{l}{$N$=Number of images.}\\
\end{tabular}
\end{table}

\begin{figure*}
  \centering
    \includegraphics[width=180mm]{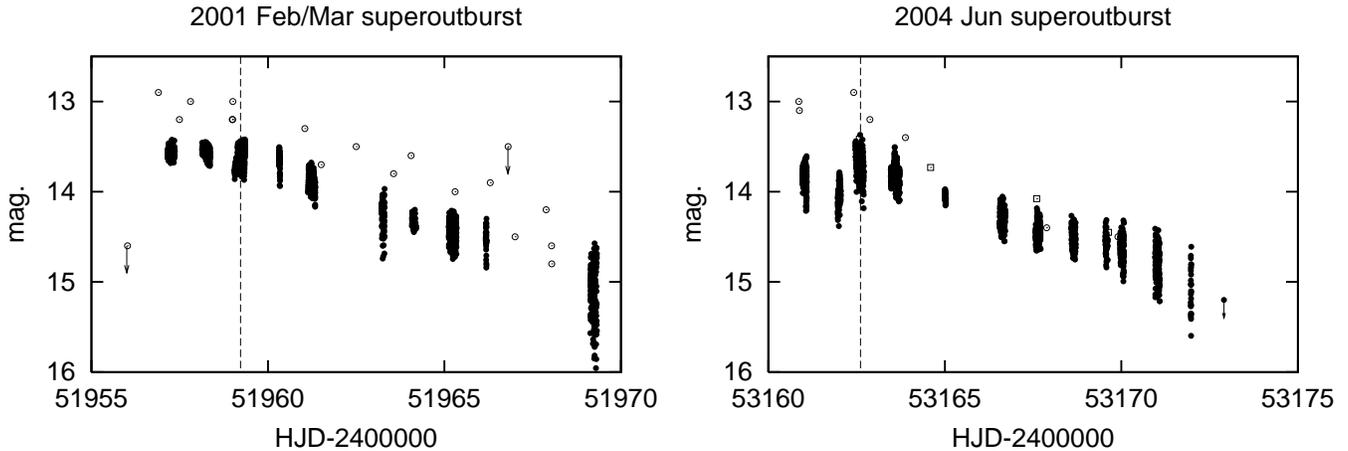}
  \caption{Light curves of the 2001 February/March (left) and 2004 June
 (right) superoutbursts.  The filled, open circles and open
 squares indicate our CCD observations, visual observations reported to
 VSNET, and observations by the ASAS-3 system, respectively
 \citep{ASAS3}.  The vertical dashed lines in each panel show times when
 superhumps have the largest amplitude.}\label{fig:lc0104} 
\end{figure*}

We conducted observational campaigns for four superoutbursts of
TV~Crv which occurred in 2001 February--March, 2002 June, 2003 May, and 2004
July, through VSNET Collaboration \citep{VSNET}.  Photometric
observations were performed with unfiltered CCD cameras attached to
30-cm class telescopes at Concepci\'on (2004), Kyoto (2001, 2002, 2003,
and 2004), Tsukuba (2001), Okayama (2002), Craigie (2003), Ellinbank
(2003), Hida (2003), and Barfold Observatory (2004).  Our observation
log is listed in Table~\ref{tab:log}.  Each image was taken with an
exposure time of $\sim 30$~s.  After correcting for the standard
de-biasing and flat fielding, we performed aperture and PSF
photometry, then obtained differential magnitudes of the object using
a neighbor comparison star UCAC2~24840990 (14.43~mag).  The constancy of
this comparison star was checked using another neighbor star
UCAC2~24840985 (14.57~mag).  In this paper, we neglect any small
differences of magnitude systems between unfiltered CCD chips used by
each observatory.  Heliocentric time corrections were applied before
the period analysis.  

\section{Results}

\begin{table*}
\caption{Observational properties of superoutbursts.}\label{tab:obssum}
\centering
\begin{tabular}{p{4cm}cccc}
\hline \hline
 & 2001 & 2002 & 2003 & 2004\\
\hline
Precursor & No & No? & No? & Yes \\
$P_{\rm SH}$ (day) & 0.065028(0.000008) & 0.064981(0.000053) &
 0.0674(0.0024) & 0.065023(0.000013) \\ 
$\dot{P}_{\rm SH}/P_{\rm SH}$ ($10^{-5}$) & 7.96(0.73) & --- & --- & $-0.32$(1.20) \\
Fading rate (mag d$^{-1}$) & 0.12(0.01) & 0.17(0.02) & 0.17(0.01) &
 0.13(0.01)\\ 
Duration (day) & 12 & 10 & 12 & 12 \\
Time interval from the last superoutburst (day) & --- & 468 & 345 & 392 \\
\hline
\end{tabular}
\end{table*}

Among the four superoutbursts, the evolution of superhumps was 
successfully detected even in early superoutburst phases during the
2001 and 2004 superoutbursts.  On the other hand, the 2002 and 2003
superoutbursts were observed rather sparsely.  We first report the
former two superoutbursts focusing on their different features, and
then shortly report the latter, poorly observed ones.  Properties of
all superoutbursts are summarized in Table~\ref{tab:obssum}.  See the
following sections for detailed information about the values in this
table.

\subsection{The 2001 and 2004 Superoutbursts}

The 2001 superoutburst was detected on February 18.392 (hereafter dates
refer to UT) at a visual
magnitude of 12.9.  Visual observations reported to VSNET
indicate that the object was fainter than 14.6~mag on February
17.517 and no pre-outburst activity is seen before February 18.  
The outburst was, hence, detected in a very early phase within one day
just after the onset of the outburst.  The first time-series CCD
observation initiated on February 18.654, about 6 hours after the visual
detection.

The 2004 superoutburst was detected on June 4.362 (UT) at a visual
magnitude of 13.0.  Observations reported to VSNET indicate that it
was fainter than 13.4~mag on May 28.399 (UT) and no pre-outburst activity
is seen before June 4.  The first time-series observation initiated on 
June 4.463 (UT), about 2 hours after the visual detection. 

The light curves of the superoutbursts in February/March 2001 and 
June 2004 are shown in Fig.~\ref{fig:lc0104}.  The most noteworthy
point in the light curves is their different behaviour during the first few
days.  While the light curve in 2001 is described with a monotonic
fading, the light curve in 2004 shows a 0.4~mag rebrightening 1.7~d
after the outburst detection.  This observation reveals that the early  
outburst was actually a precursor of the late genuine supermaximum.  In
conjunction with the close monitoring of the object, we conclude that
no precursor event was associated with the 2001 superoutburst.

\begin{figure*}
  \centering
    \includegraphics[width=180mm]{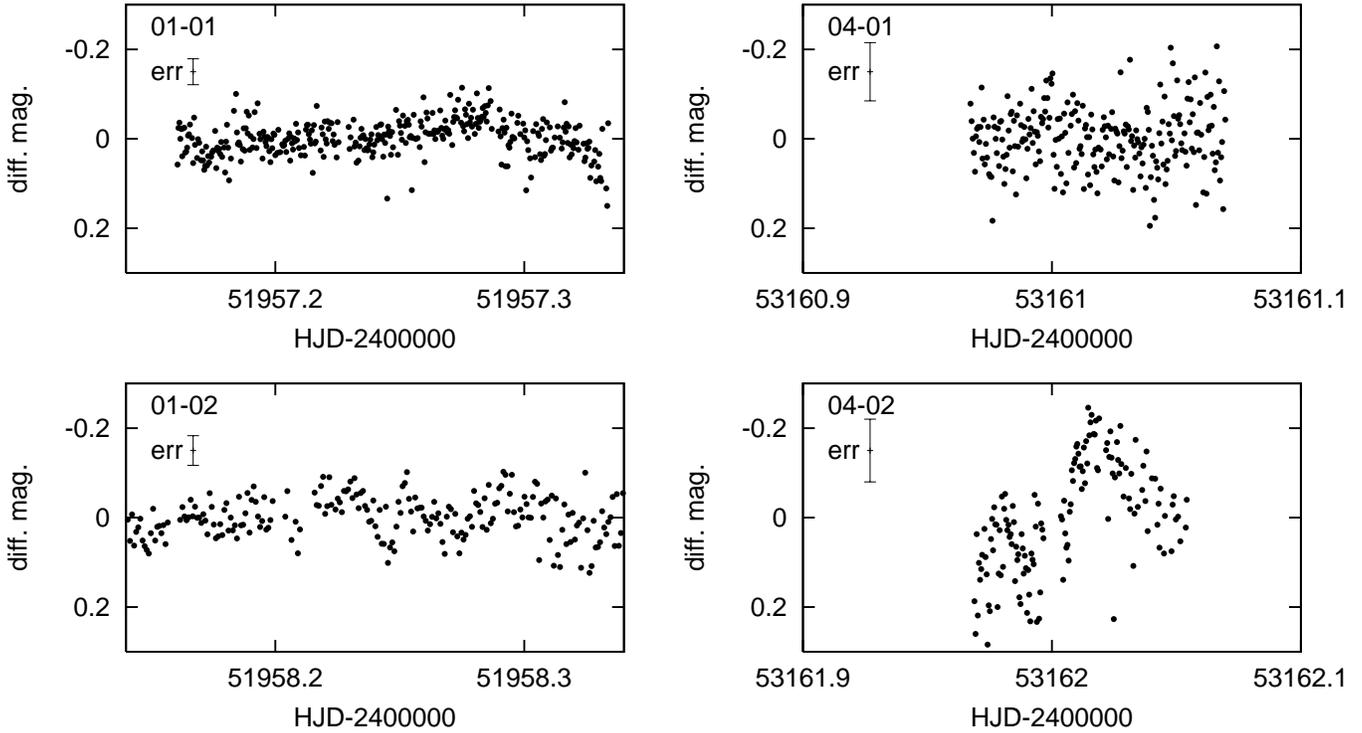}
  \caption{Light curves during early phases of superoutbursts in 2001
 and 2004.  The abscissa and ordinate denote the time in HJD and the
 differential magnitudes, respectively.  The magnitude system is
 normalized by subtracting average magnitudes of each panel.  We
 indicate the run ID number (Table~\ref{tab:log}) and typical errors in
 each panel.}\label{fig:pre0104}
\end{figure*} 

We succeeded in obtaining time-series data during the early phase of
the superoutbursts, which are shown in Fig.~\ref{fig:pre0104}.  We
also show the observation IDs (see Table~\ref{tab:log}) and typical
errors in each panel.  As can be seen in Fig.~\ref{fig:pre0104}, no
superhump-like modulation appears except for the ``04-02'', in which a
0.3-mag hump is detected.  The ``04-02'' run lasted 2.05~hr which well
covers an orbital period of TV~Crv.  Throughout this run, the object is
on a rapid brightening trend at a rate of 2.6~mag~d$^{-1}$.  The hump is
superimposed on this brightening trend.  This indicates that the
temporary fading from the precursor had already been terminated, and
then started brightening to the supermaximum during the ``04-02'' run.    

The other panels of the ``01-01'', ``01-02'', and ``04-01'' in
Fig.~\ref{fig:pre0104} show modulations with rather small amplitudes
($\sim 0.1\;{\rm mag}$) and long timescales.  No periodic signal is
detected in these runs with our Fourier analysis in the period range of
10~s--0.1~d.  On the other hand, we note that possible 
0.1--0.2~mag amplitude short-term fluctuations with timescale of $\sim
10\;{\rm min}$ can be seen in the ``01-02'' run.

We detected superhumps after this early phase.  In the
case of the 2001 superoutburst, fully grown superhumps appeared on
JD~2451959 (the ``01-04'' run).  In the case of the 2004 
superoutburst, on the other hand, the supermaximum coincides with
the apparition  of superhumps with the largest amplitude of $\sim 0.4\;{\rm
mag}$ (the ``04-03'' run).  

\begin{figure*}
  \centering
    \includegraphics[width=180mm]{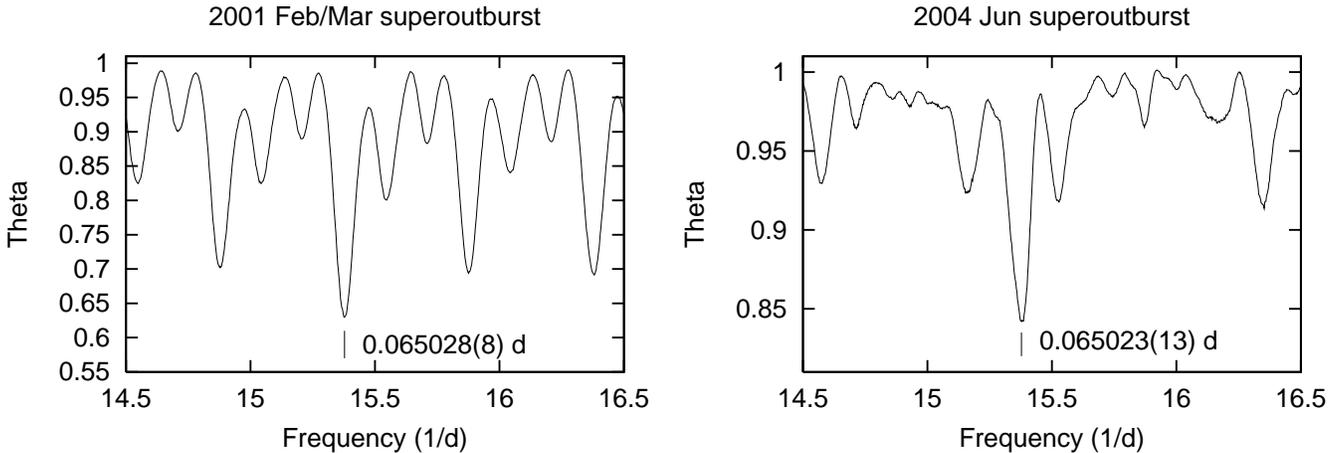}
  \caption{Frequency-$\Theta$ diagrams for the 2001 (left) and 2004 (right)
 superoutbursts calculated by the PDM method.  }\label{fig:pdm0104}
\end{figure*}

\begin{figure*}
  \centering
    \includegraphics[width=120mm]{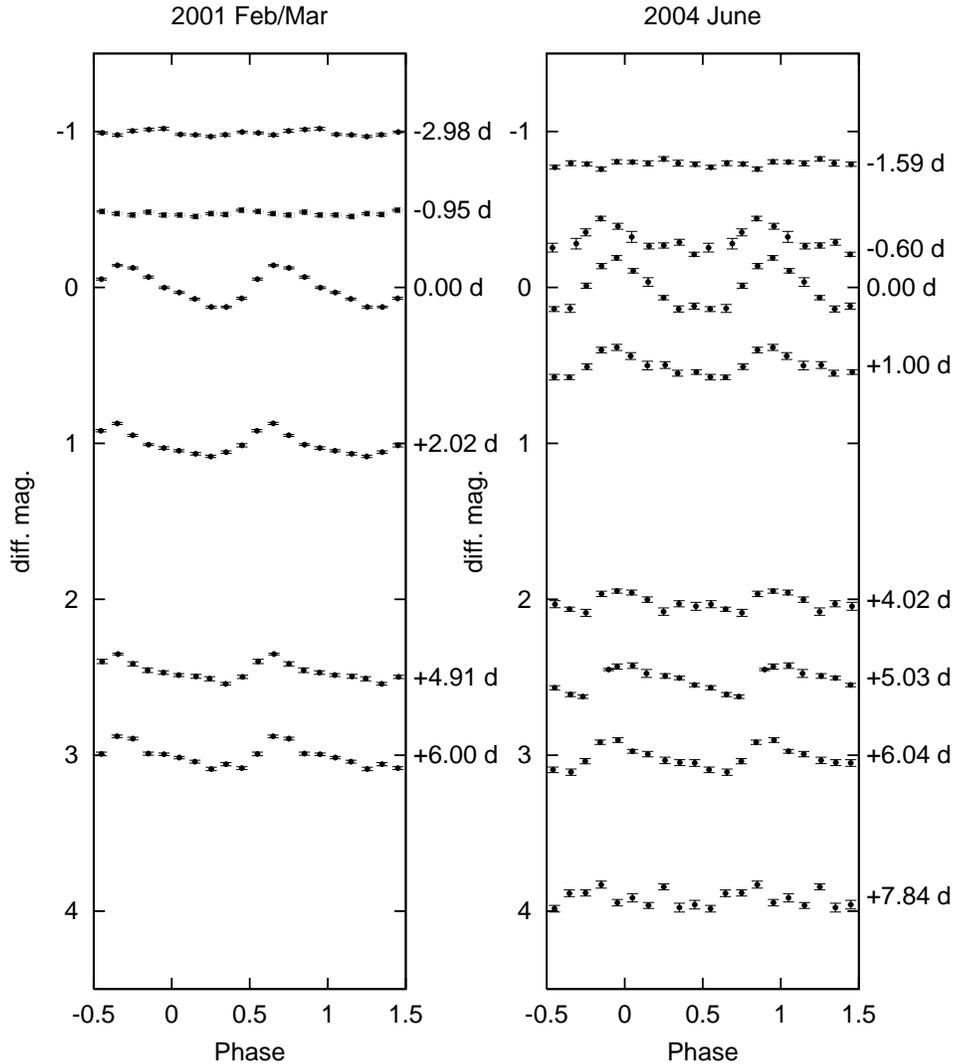}
  \caption{Superhump evolution during the 2001 (left) and 2004 (right)
 superoutbursts.  The abscissa and ordinate denote the superhump phase
 and the differential magnitude, respectively.  The phase is calculated
 with a superhump period of 0.065028~d and an arbitrary epoch.  The
 differential magnitudes are normalized by each average magnitude, and
 are sorted with observation times which are indicated on the right
 vertical axis of each panel.  See the text for detailed
 information.}\label{fig:hump0104}  
\end{figure*}

\begin{figure*}
  \centering
    \includegraphics[width=180mm]{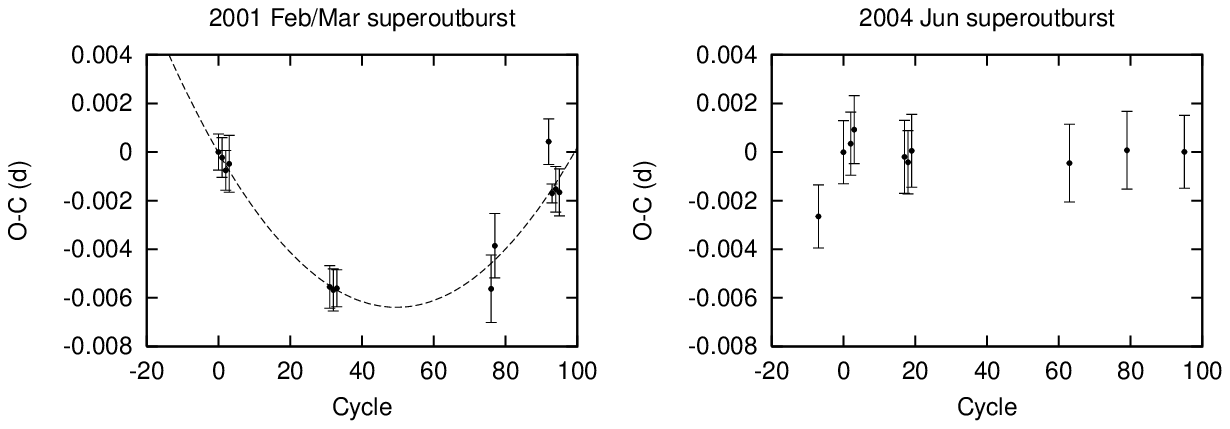}
  \caption{$O-C$ diagrams of superhumps during the 2001 (left) and 2004
 (right) superoutbursts.  The abscissa and ordinate denote the cycle and
 the $O-C$ in day, respectively.  The dashed line in the left panel is
 the best fitted quadratic curve for the $O-C$ in
 2001.}\label{fig:o-c0104} 
\end{figure*}

A period analysis with the PDM method \citep{PDM} was performed after
linear trends were subtracted from the light curves.  We used
light curves between JD~2451959.0 and 2451966.2 for the 2001
superoutburst and between JD~2453162.4 and 2453171.1 for
the 2004 superoutburst.  The samples for the 2001 and 2004
superoutbursts contain 1830 and 1692 photometric points,
respectively.  The PDM analysis yielded the frequency--$\Theta$ diagram
shown in Fig.~\ref{fig:pdm0104}.  The superhump periods are calculated to
be $0.065028\pm 0.000008$~d (2001) and $0.065023\pm 0.000013$~d
(2004).  These are in agreement each other and also in agreement with
$P_{\rm SH}$ reported in \citet{how96tvcrv} ($0.0650\pm 0.0008\;{\rm
d}$).  Since the error of $P_{\rm SH}$ is smaller in 2001 than that in
2004, we adopt $P_{\rm SH}$ of TV~Crv to be $0.065028\pm 0.000008$~d in
this paper.  According to \citet{wou03tvcrv}, the orbital period of
TV~Crv is $0.06288\pm 0.00013$~d, which yields a superhump period
excess $\varepsilon=0.0342\pm 0.0021$.  The 3.4\% superhump excess is
relatively large for short-period SU~UMa systems \citep{pat03cvs}.

Fig.~\ref{fig:hump0104} shows the evolution of the superhumps from the
early phase including the precursor to the end of the superoutburst
plateau.  All light curves are folded with $P_{\rm SH}=0.065028$~d 
and an arbitrary epoch.  The abscissa and ordinate denote the
phase and the differential magnitude, respectively.  We calculated
center times of each run and show them in the figure.
We set the origin of the times at the ``01-04'' and ``04-03'' runs, in
which superhumps had the largest amplitude.  The differential magnitudes
are normalized by each average magnitude, and are shifted by constants
proportional to the times of each run in order to clearly compare two
sequences.  The hump just before the supermaximum on 2004 has
a peak phase roughly the same as those of later superhumps.  It strongly
indicates that the hump is actually a  superhump, growing to the
supermaximum, as observed in T~Leo \citep{kat97tleo}, V436~Cen
\citep{sem80v436cen}, and GO~Com \citep{ima04gocom}.  As can be seen
from both panels, the amplitude of superhumps decreased in a
few days, then kept 0.2-mag peak-to-peak amplitudes for 6~days.  The
2001 and 2004 superoutbursts, thus, have quite similar characteristics
regarding the evolution of superhump amplitudes.  

We determined peak times of superhumps by taking cross-correlation
between the light curve and average profiles of superhumps.
With determined peaks and $P_{\rm SH}$ of 0.065028~d, we
calculate the $O-C$ of the superhump maximum timings, which is shown in
Fig.~\ref{fig:o-c0104}.  There is an obvious difference between the
$O-C$ in the 2001 and 2004 superoutbursts.  The $O-C$ clearly
indicates an increase of $P_{\rm SH}$ with time in the case of the
2001 superoutburst.  A quadratic fit to the $O-C$ yields a period
derivative of $\dot{P}_{\rm SH}/P_{\rm SH}=7.96\pm 0.73\times 10^{-5}$.  On
the other hand, the $O-C$ is almost constant, in other words, $P_{\rm
SH}$ was stable during the 2004 superoutburst.  A quadratic fit yields 
$\dot{P}_{\rm SH}/P_{\rm SH}=-0.32\pm 1.20\times 10^{-6}$.  This result
indicates that the superhumps in 2004 superoutburst have quite small
$\dot{P}_{\rm SH}/P_{\rm SH}$ compared with other systems
\citep{kat03v877arakktelpucma}. 

We note that there is a slight phase shift at the hump just before the
supermaximum in 2004 superoutburst, as shown in the right panel of
Fig.~\ref{fig:o-c0104}.  The slight phase shift in the early stage
implies that superhumps evolved with a rapid period change just before
the supermaximum.  Similar rapid period changes during very early phases
are also known in T~Leo \citep{kat97tleo}, V1028~Cyg
\citep{bab00v1028cyg}, and  XZ~Eri \citep{uem04xzeri}. 

\subsection{The 2002 Superoutburst}

\begin{figure}
  \centering
    \includegraphics[width=88mm]{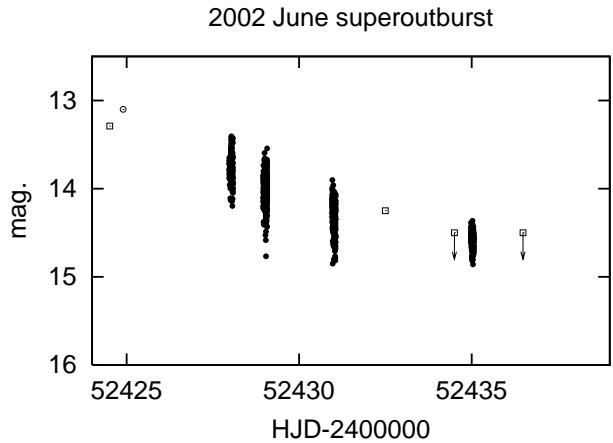}
  \caption{Light curve of the superoutburst in 2002 June.  The symbols
 are the same as in Fig.~\ref{fig:lc0104}.}\label{fig:lc2002}
\end{figure}

\begin{figure}
  \centering
    \includegraphics[width=88mm]{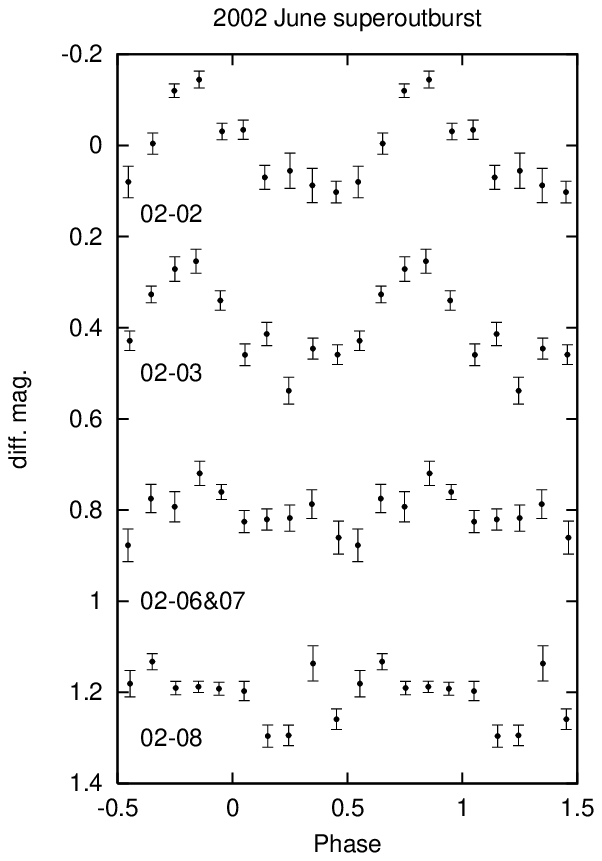}
  \caption{Superhump evolution during the 2002 superoutburst.  The
 symbols in the figure are the same as in
 Fig.~\ref{fig:hump0104}.}\label{fig:hump2002} 
\end{figure}

\begin{figure}
  \centering
    \includegraphics[width=88mm]{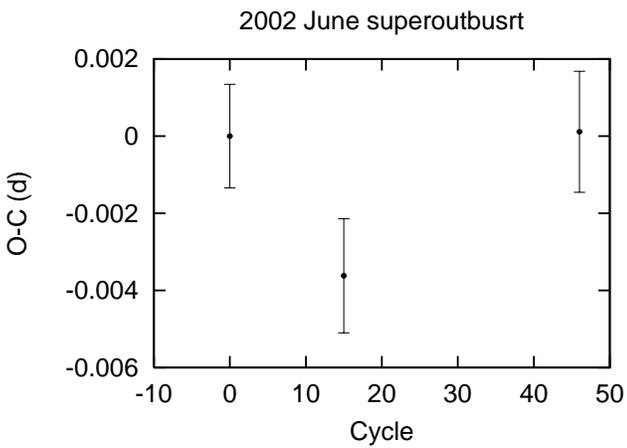}
  \caption{$O-C$ diagram of superhumps during the 2002 superoutburst.
 The symbols in the figure are the same as in
 Fig.~\ref{fig:o-c0104}.}\label{fig:o-c02}
\end{figure}

The 2002 superoutburst was first detected on May 30.399 (UT) at a visual
magnitude of 13.1~mag.  The ASAS-3 system records an earlier
detection of the outburst on May 30.009 (UT) and a negative detection on
May 21.048 (UT) \citep{ASAS3}.  Unfortunately, there is no time-series
data just after the outburst detection.  The first run (the ``02-01''
run in Table~\ref{tab:log}) initiated at June 2.476 (UT).  The light curve
of the superoutburst is shown in Fig.~\ref{fig:lc2002}.  The ``02-01''
run detected superhumps, which establish that this outburst is a
superoutburst.  Profiles of superhumps during this superoutburst are
shown in Fig.~\ref{fig:hump2002}.  Fig.~\ref{fig:o-c02} is the $O-C$
diagram of superhumps.  While it contains only three points, this figure
apparently implies a period increase of superhumps, as observed in the
2001 superoutburst. 

\subsection{The 2003 Superoutburst}

\begin{figure}
  \centering
    \includegraphics[width=88mm]{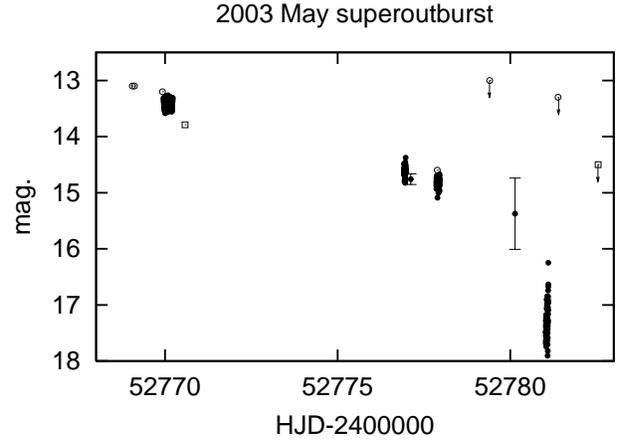}
  \caption{Light curve of the superoutburst in 2003 May.  The symbols
 in the figure are the same as in Fig.~\ref{fig:lc0104}.}\label{fig:lc2003}
\end{figure}

\begin{figure}
  \centering
    \includegraphics[width=88mm]{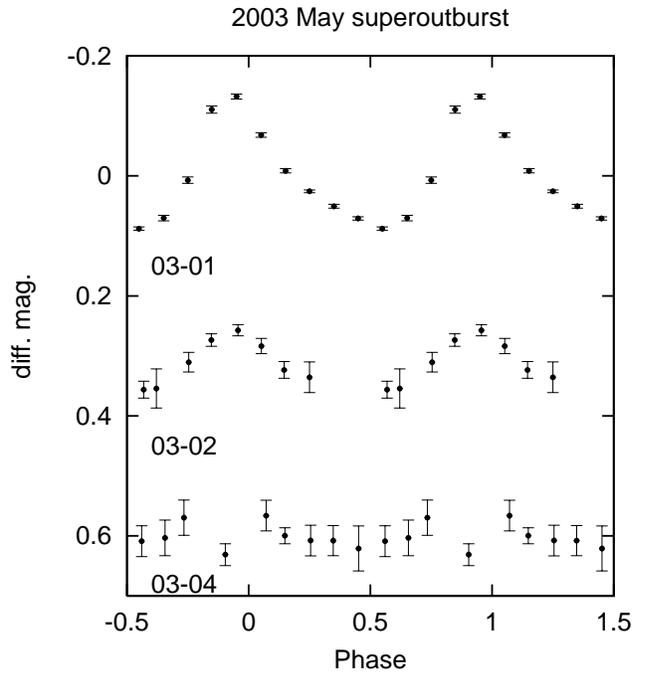}
  \caption{Superhump evolution during the 2003 superoutburst.  The
 symbols in the figure are the same as in
 Fig.~\ref{fig:hump0104}.}\label{fig:hump2003}
\end{figure}

The 2003 superoutburst was discovered by a visual observation on May
9.546 (UT) at 13.1~mag.  The latest negative visual observation had been 
reported on May 6.412 (UT) (fainter than 14.6~mag), three days before
the outburst detection.  The first time-series observation initiated
at May 10.458 (UT), about one day after the outburst detection.
Considering the rapid evolution during the precursor in the 2001
superoutburst, we cannot exclude the possibility that the 
2003 superoutburst had a precursor between May 6 and 9.  The light
curve of this outburst is shown in Fig.~\ref{fig:lc2003}.  The
first run ``03-01'' clearly detects fully grown superhumps, as shown in
Fig.~\ref{fig:hump2003}, which reveal that it is another
superoutburst.  Due to the lack of enough observations, we cannot find
any hints of significant period changes of superhumps.

\section{Implication for the TTI model}

The observational properties of the four superoutbursts are summarized in
Table~\ref{tab:obssum}.  TV~Crv is one of the typical short orbital period SU~UMa-type dwarf
novae.  Its supercycle
is calculated to be $402\pm 51$~d from the three time-intervals of
superoutbursts listed in Table~\ref{tab:obssum}.  This supercycle is
also a typical value for SU~UMa stars.  A noteworthy feature of TV~Crv
is its superhump excess (3.4\%), which is relatively large for
short-period systems, but not extraordinary \citep{pat03cvs}.  It is
well known that the superhump period excess is related to the
superhump period \citep{pat03cvs}.  From the theoretical point of view,
this can be understood since the precession velocity of the eccentric
disk depends on the disk radius and the mass ratio of binary systems.  
The superhump period excess, $\varepsilon$, can be expressed as
\citep{osa85SHexcess}; 
\begin{eqnarray}
\varepsilon= {3 \over 4}{q \over \sqrt{1+q}} \left( {r_{\rm d} \over a}
\right)^{3/2}.
\end{eqnarray}
Assuming a certain disk radius at which the
tidal mode is excited, one can describe the superhump period excess as a
function of the superhump period \citep{min92BHXNSH}.  The large superhump
excess of TV~Crv, therefore, implies a relatively large mass ratio
among short-period SU~UMa stars.  The empirical relationship in
\citet{pat01SH} yields the mass ratio to be $q=0.16\pm 0.01$ for
TV~Crv.  On the other hand, it is possible that the large superhump
excess is partly caused by an unusually large disk radius.  The large
mass ratio of TV Crv should be confirmed by spectroscopic observations
in future.  

Our observations reveal that TV~Crv experiences two types of
superoutbursts, that is, one with a precursor and the other without a
precursor.  Similar morphology studies of superoutburst light curves
had been performed for VW~Hyi, which also shows the two types of
superoutbursts \citep{bat77vwhyi,mar79superoutburst}.  VW~Hyi is 
a typical SU~UMa-type dwarf novae having a relatively long orbital
period of 0.074271~d \citep{DownesCVatlas3}.  Our observations of TV~Crv
are the first to show that those two types of superoutbursts appear even in short orbital period
systems.

To explain the behaviour of VW~Hyi, \citet{osa03DNoutburst} propose the
refined TTI model, in which the types of superoutburst depend on the
maximum radius of the accretion disk. When the accretion disk reaches
the tidal truncation radius, the dammed matter prevents the disk from a
propagation of a cooling wave, leading to a superoutburst without a
precursor.  In this view, a large mass ratio is required for a system to
achieve the situation that the tidal truncation radius lies just beyond
the 3:1 resonance radius.  On the other hand, when the disk fails to
reach the tidal truncation radius, a rapid fading initiates.  This
fading is terminated, and the object rebrightens to a supermaximum due
to a growth of the tidal dissipation.  In this case, a large mass ratio
is also required for a rapid growth of the tidal dissipation before the
object returns to quiescence.  VW~Hyi has a superhump excess of 3.9\%
\citep{vaname87vwhyi}, which yields a mass ratio $q=0.18$ from the
empirical relationship in \citet{pat01SH}.  \citet{tap03CVatlas}
reported $q\sim 0.14$ for VW~Hyi based on their spectroscopic
observations.  The mass ratio of TV~Crv is possibly close to that of
VW~Hyi rather than those of ordinary short period SU~UMa stars
\citep{pat01SH}. 

Although TV~Crv is a short period system, we propose that it has a
relatively large mass ratio.  According to \citet{osa03DNoutburst}, a
system having a large mass ratio ($q\sim 0.2$) can experience the two
types of superoutburst.  The behaviour of TV~Crv can, therefore, be
explained by the refined TTI model, furthermore, it possibly provides
evidence that the mass ratio plays a key role in the morphology of
superoutburst light curve.  

\section{Presence of a precursor and superhump evolution}

The most important and unforeseen finding in our observation is 
that the $\dot{P}_{\rm SH}/P_{\rm SH}$ can be variable in distinct
superoutbursts in one system.  This is clearly shown in
Table~\ref{tab:obssum}; a positive $\dot{P}_{\rm SH}/P_{\rm SH}$ in the
2001 superoutburst and an almost constant $P_{\rm SH}$ in the 2004
superoutburst.  Except for the difference in $\dot{P}_{\rm SH}/P_{\rm
SH}$, another observational difference between these two superoutbursts
is the presence or absence of the precursor.  There was no precursor in
the 2001 superoutburst, while a clear precursor was observed in 2004.
Our observation hence indicates that the $\dot{P}_{\rm SH}/P_{\rm SH}$
is related to the precursor phenomenon.

As mentioned above, the TTI model suggests that the appearance of the
precursor depends on whether the disk reaches the tidal truncation
radius or not.  Based on this idea, at the time when superhumps are
fully grown, the disk size should be different in the two types of 
superoutbursts.  In the case of the precursor-main type outburst, the
disk size is around the 3:1 resonance radius at  supermaximum.  On the other
hand, in the case of the superoutburst without a precursor, the hot disk
can remain larger than the 3:1 resonance radius due to the dammed matter at the
tidal truncation radius.  The accretion disk can, hence, have a
relatively large amount of gas beyond the 3:1 resonance radius even a few
days after the supermaximum when superhumps are fully grown.  We
therefore propose that the $\dot{P}_{\rm SH}/P_{\rm SH}$ is related to
the amount of the gas around and beyond the 3:1 resonance radius.  

We now present an idea how the disk size actually affects the
eccentric disk evolution.  We first consider the standard picture of
the eccentric disk evolution.  In an early phase of outburst, the
rapid excitation of the eccentric mode stops when the angular momentum
removal by the tidal dissipation is balanced with the input angular
momentum transfered from the inner region.  In the case of the
precursor-main type outburst, then, the accretion disk shrinks below the
3:1 resonance radius at that time \citep{whi94SH}.  The eccentricity
wave can only propagate inward, since the tidal mode is no longer
excited.  In the case of the superoutburst without a precursor, on the
other hand, we can expect a large amount of gas over the 3:1 resonance
radius at that time.  We conjecture that the eccentric mode can keep
excited because the disk radius presumably remains larger than the 3:1
resonance radius.  The positive $\dot{P}_{\rm SH}/P_{\rm SH}$ can be
explained by a gradual outward propagation of the eccentricity wave.  

It is, however, unclear whether the outward propagation is possible only
with the large disk.  The outward propagation essentially requires an
additional input of angular momentum from an inner region.  It might be
possible that the gas in an inner region may be swept up, then give
additional angular momentum into the outermost area of the eccentricity
wave.  This additional supply of angular momentum would enable to keep
the disk size large and the continuous excitation of the eccentric mode.

\citet{ole03ksuma} propose that the $\dot{P}_{\rm SH}/P_{\rm SH}$ is
negative at the beginning and the end of the superoutburst, but positive in
the middle phase for several SU~UMa-type dwarf novae.  Based on our
scenario, the duration of the positive $\dot{P}_{\rm SH}/P_{\rm SH}$
depends on the amount of the gas which enables the continuous
excitation of the eccentric mode.  The transition from a positive
$\dot{P}_{\rm SH}/P_{\rm SH}$ to a negative one may be explained by the 
depletion of the gas.  

The above discussion is summarized in the following two ideas: i) At
the time when superhumps are fully grown, the accretion disk remains
larger in the superoutburst without the precursor than in the
precursor-main type superoutburst.  ii) Even after that, the eccentric 
mode keeps excited through a superoutburst.  These ideas should be
tested by hydrodynamical simulations.  

\section{Discussion}

\begin{figure}
  \centering
    \includegraphics[width=88mm]{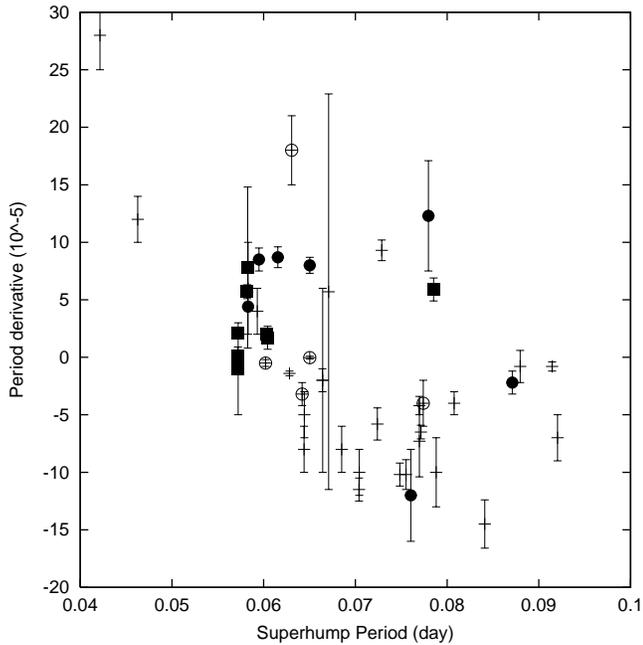}
  \caption{The superhump period derivative against the superhump period
 for the SU~UMa-type dwarf novae listed in Table~\ref{tab:ref}.  The
 open circles indicate type A superoutbursts which have a precursor.
 The filled circles indicate type B superoutbursts in which a delay of
 superhump growth is observed.  The filled squares indicate WZ~Sge-type
 dwarf novae.  The other points indicated by the crosses are objects
 whose outburst types are unknown.  The figure focuses on objects whose
 outburst types are known.  We, hence, omit three unknown-type dwarf
 novae, KK~Tel, MN~Dra, (exceptionally large period derivatives) and
 TU~Men (a long superhump period) in this figure.  We only show positive
 values of period derivatives for KS~UMa and TT~Boo, in which changes of
 the period derivative have been observed.}\label{fig:ref}
\end{figure}

We revealed that the $\dot{P}_{\rm SH}/P_{\rm SH}$
is variable in distinct superoutbursts for TV Corvi.  This result should be
confirmed by observations of other sources in future because we now
have no data of variations of the $\dot{P}_{\rm SH}/P_{\rm SH}$ against 
different types of superoutburst in other sources.  On the other hand, 
it is valuable to investigate the relationship of $\dot{P}_{\rm
SH}/P_{\rm SH}$ and the type of superoutburst in known systems.  To
perform this, we collected the sample of 40 dwarf novae and one X-ray
binary whose $\dot{P}_{\rm SH}/P_{\rm SH}$ is published, as listed in 
Table~\ref{tab:ref}.  We now classify the morphology of superoutburst
light curve into two types, that is, the type ``A'' and type ``B''.  The
type A is defined by the detection of a precursor, in other words, the 
precursor-main type superoutburst.  On the other hand, the type B is
defined by the detection of a delay of the superhump growth after a
supermaximum.  In our sample listed in Table~\ref{tab:ref}, we find 6
and 7 cases for the type A and B, respectively.  There is no system
having both features.  WZ~Sge-type dwarf novae are indicated by ``WZ''
in Table~\ref{tab:ref}  because their superhump evolution is peculiar;
they have an early hump era followed by an ordinary superhump era
\citep{kat96alcom}.  The sample in Table~\ref{tab:ref} has 8 cases for 5
WZ~Sge stars.  Types of superoutburst are unclear in the other
29 cases due to the lack of enough observations during early phases of
superoutbursts.  The $\dot{P}_{\rm SH}/P_{\rm SH}$ are shown against
$P_{\rm SH}$ in Fig.~\ref{fig:ref}.  

As mentioned above, WZ~Sge-type systems tend to show positive 
$\dot{P}_{\rm SH}/P_{\rm SH}$ as indicated by filled squares in
Fig.~\ref{fig:ref}.  In these systems, their long recurrence
time and the lack of normal outburst lead to a huge amount of
accumulated gas compared with ordinary SU~UMa systems.  At the onset
of their outburst, the accretion disk, hence, violently expands beyond
the 3:1 resonance radius.  The large disk in WZ~Sge stars may partly be
due to a continuous expansion of their quiescent disks, as proposed
in \citet{min98wzsge}.  This situation in WZ~Sge systems is similar to
the type B outburst in TV~Crv discussed in the last section.  
\citet{kat04egcnc} propose a scenario analogous to that described in the last
section for positive $\dot{P}_{\rm SH}/P_{\rm SH}$ in WZ~Sge-type dwarf
novae.  The difference between WZ~Sge systems and TV~Crv is the
mechanism to generate a large disk over the 3:1 resonance radius.  
In the case of WZ~Sge systems, \citet{osa03DNoutburst} propose that
the large disk is maintained by the strong tidal removal of angular
momentum at the 2:1 resonance radius.  The disk can reach the 2:1 resonance
radius because of the large amount of accumulated matter.  In the case
of the type B outbursts of TV~Crv, the large disk is maintained at the
tidal truncation radius.  This is due to a high mass ratio leading to
the tidal truncation radius just beyond the 3:1 resonance radius.

We can therefore consider that a similar physical condition appears in
the type B superoutbursts and the WZ~Sge-type superoutbursts, in terms of 
the superhump evolution.  In Fig.~\ref{fig:ref}, we show these
objects as filled symbols (squares for WZ~Sge stars and circles for
the type B) and the type A superoutburst as open circles.  We can
see a tendency that the B- and WZ-types generally have larger 
$\dot{P}_{\rm SH}/P_{\rm SH}$, as expected from our scenario.  This
figure, however, also show the presence of two exceptions breaking the
tendency (GO~Com and V1251~Cyg).  The nature of these objects is 
an open issue.  We need to obtain their $\dot{P}_{\rm SH}/P_{\rm SH}$
in another superoutburst to investigate  their possible
variations.

The two systems having the shortest $P_{\rm SH}$ in
Fig.~\ref{fig:ref} are V485~Cen and EI~Psc.  While they have 
ultra-short orbital periods, their secondaries are relatively massive
\citep{aug93v485cen,tho02j2329}.  The superhump period excess and mass
ratio of EI~Psc are $\varepsilon=0.040$ and $q=0.19$, respectively,
which are actually larger than those of TV~Crv and VW~Hyi
\citep{uem02j2329letter}.  According to the refined TTI model, their
accretion disks can reach the tidal truncation radius and remain active
in the eccentric mode through a superoutburst.  Their high
$\dot{P}_{\rm SH}/P_{\rm SH}$ can, hence, be naturally explained with
our scenario.  Observations of the onset of their superoutbursts are
encouraged to reveal the type of them.  

The only X-ray binary in table 3, XTE~J1118+480, is a black hole X-ray binary (BHXB) having a quite low
mass ratio $q=0.05$ \citep{wag01j1118}.  This is a unique object in
the point that the $\dot{P}_{\rm SH}/P_{\rm SH}$ is significantly
determined in BHXBs.  Although the low mass ratio implies a situation
similar to WZ~Sge-type dwarf novae, its $\dot{P}_{\rm SH}/P_{\rm SH}$
is slightly, but significantly negative as listed in
Table~\ref{tab:ref}.  On the other hand, its main outburst has a
precursor, which is reminiscent of the precursor-main type
superoutburst in SU~UMa systems \citep{kuu01j1118}.  The accretion
disk radius was probably just around the 3:1 resonance radius at the
``supermaximum'' of XTE~J1118+480.  This rather small disk may
cause the inward propagation of an eccentricity wave in this low-$q$
system.

\section{Summary}

Our findings through observations of four superoutbursts of TV~Crv are
summarized below:\\
i) We accurately determined the superhump period to be $0.065028\pm
0.000008$~d. \\
ii) In conjunction with the orbital period in \citet{wou03tvcrv}, the
superhump period yields a high superhump period excess of $0.0342\pm
0.0021$.  This implies that TV~Crv has a relatively large mass ratio
compared with other short-period SU~UMa systems.  Using the empirical
relationship for the superhump mass ratio in \citet{pat01SH}, the mass
ratio of TV~Crv is estimated to be $q=0.16\pm 0.01$.\\
iii) TV~Crv experiences two types of
superoutbursts; one with a precursor and the other without.  This
behaviour can be interpreted with the refined thermal-tidal instability
model if TV~Crv has a relatively large mass ratio in spite of its short
orbital period.\\
iv) We show that the superhump period derivative is variable in
distinct superoutbursts.  The difference is apparently related to the
presence/absence of a precursor.\\  
v) We propose that the eccentric mode keeps excited when the accretion
disk remains larger than the 3:1 resonance radius.  This scenario can
explain the behaviour of TV~Crv, and furthermore be consistent with
systematically large period derivatives in superoutbursts without a
precursor.\\

We greatly appreciate valuable comments made by Dr. Shin Mineshige. 
This work is supported by the Grant-in-Aid for the 21st
Century COE ``Center for Diversity and Universality in Physics'' from the
Ministry of Education, Culture, Sports, Science and Technology (MEXT) of
Japan.  RM acknowledges grant Fondecyt 1030707.  PN acknowledges the
Curry Foundation and the AAVSO.  This work is partly supported by a
grant-in aid from the Japanese Ministry of Education, Culture, Sports,
Science and Technology (No.s. 13640239, 15037205).  Part of this work is
supported by a Research Fellowship of the Japan Society for the
Promotion of Science for Young Scientists.

\bibliographystyle{aa}

\begin{longtable}{llccl}
  \caption{Superhump period derivative and the type of
 superoutbursts.}\label{tab:ref} \\
\hline \hline
Object & $P_{\rm SH}$ & $\dot{P}_{\rm SH}/P_{\rm SH}$ & Type & Ref.\\
 & (day) & ($10^{-5}$) & & \\
\hline
\endfirsthead
\caption{continued.}\\
\hline \hline
Object & $P_{\rm SH}$ & $\dot{P}_{\rm SH}/P_{\rm SH}$ & Type & Ref.\\
 & (day) & ($10^{-5}$) & & \\
\hline
\endhead
\hline
\endfoot
\hline
\multicolumn{5}{p{180mm}}{References: 1.\citet{ole97v485cen}, 2.\citet{uem02j2329},
 3.\citet{kuu02wzsge}, 4.\citet{ish02wzsgeletter}, 5.\citet{nog97alcom},
 6.\citet{kat01hvvir}, 7.\citet{ish03hvvir}, 8.\citet{nog98swuma},
 9.\citet{kat01wxcet}, 10.\citet{kat97tleo}, 11.\citet{kat97egcnc},
 12.\citet{kat04egcnc}, 13.\citet{ima04gocom}, 14.\citet{bab00v1028cyg},
 15.\citet{uem04xzeri}, 16.\citet{pat95v1159ori}, 17.\citet{pat93vyaqr},
 18.\citet{sch86oycar}, 19.this work, 20.\citet{kat99cthya},
 21.\citet{nog03dmlyr}, 22.\citet{nog97sxlmi}, 23.\citet{ole03ksuma},
 24.\citet{kat96rzsge}, 25.\citet{sem97rzsge}, 26.\citet{har95cyuma},
 27.\citet{nog04vwcrb}, 28.\citet{kat03nsv10934mmscoabnorcal86}, 
 29.\citet{kat02cccnc}, 30.\citet{kat95v1251cyg}, 31.\citet{nog04qwser},
 32.\citet{hae79lateSH}, 33.\citet{kuu91zcha}, 34.\citet{ole04ttboo}, 
 35.\citet{ish01rzleo},
 36.\citet{uda90suuma}, 37.\citet{kat98hsvir},
 38.\citet{kat03v877arakktelpucma}, 39.\citet{kat02efpeg},
 40.\citet{kat03bfara}, 41.\citet{kat93v344lyr}, 42.\citet{pat79SH},
 43.\citet{uem01v725aql}, 44.\citet{nog03var73dra}, 45.\citet{sto84sp},
 46.\citet{uem02j1118}}
\endlastfoot
V485 Cen    & 0.04216 &  28(3)     & -- & 1\\
EI Psc      & 0.04627 &  12(2)     & -- & 2\\
WZ Sge(1978)& 0.05722 & $-$1(4)      & WZ & 3\\
WZ Sge(2001)& 0.05719 &  0.1(0.8)  & WZ & 4\\
AL Com(1995)& 0.0572  &  2.1(0.3)  & WZ & 5\\
HV Vir      & 0.05820 &  5.7(0.6)  & WZ & 6\\
HV Vir(2002)& 0.05826 &  7.8(7)    & WZ & 7\\
SW UMa(1991)& 0.0583  &  6(4)      & -- & 6\\
SW UMa(1996)& 0.0583  &  4.4(0.4)  & B  & 8\\
WX Cet(1996)& 0.0593  &  4(2)      & -- & 6\\
WX Cet(1998)& 0.05949 &  8.5(1.0)  & B  & 9\\
T Leo       & 0.0602  & $-$0.5(0.3)  & A  & 10\\
EG Cnc      & 0.06038 &  2.0(0.4)  & WZ & 11\\
EG Cnc      & 0.06043 &  1.7(1)    & WZ & 12\\
GO Com      & 0.06306 &  18(3)     & A  & 13\\
V1028 Cyg   & 0.06154 &  8.7(0.9)  & B  & 14\\
XZ Eri      & 0.06281 & $-$1.4(0.2)  & -- & 15\\
V1159 Ori   & 0.0642  & $-$3.2(1)    & A  & 16\\
VY Aqr      & 0.0644  & $-$8(2)      & -- & 17\\
OY Car      & 0.06443 & $-$5(2)      & -- & 18\\
TV Crv(2001)& 0.06503 &  8.0(0.7)  & B  & 19\\
TV Crv(2004)& 0.06502 & $-$0.03(0.12)& A  & 19\\
UV Per      & 0.06641 & $-$2.0(1)    & -- & 6\\
CT Hya      & 0.06643 & $-$2(8)      & -- & 20\\
DM Lyr      & 0.06709 &  5.7(17.2) & -- & 21\\
SX LMi      & 0.0685  & $-$8(2)      & -- & 22\\
KS UMa(2003 early) & 0.07009 & $-$21(8) & -- & 23\\
KS UMa(2003 late)  & 0.07009 & 21(12) & -- & 23\\
RZ Sge(1994)& 0.07042 & $-$10(2)     & -- & 24\\
RZ Sge(1996)& 0.07039 & $-$11.5(1)   & -- & 25\\
CY UMa      & 0.0724  & $-$5.8(1.4)  & -- & 26\\
VW Crb      & 0.07287 &  9.3(0.9)  & -- & 27\\
NSV 10934   & 0.07485 & $-$10.2(1.0) & -- & 28\\
CC Cnc      & 0.07552 & $-$10.2(1.3) & -- & 29\\
V1251 Cyg   & 0.07604 & $-$12(4)     & B  & 30\\
QW Ser(2000)& 0.07698 & $-$4.2(0.8)  & -- & 31\\
QW Ser(2002)& 0.07697 & $-$7.3(3.1)  & -- & 31\\
VW Hyi      & 0.07714 & $-$6.5(0.6)  & -- & 32\\
Z Cha       & 0.07740 & $-$4(2)      & A  & 33\\
TT Boo(2004 early) & 0.07796 & $-$52.3(1.3) & B & 34\\
TT Boo(2004 middle) & 0.07796 & 12.3(4.8) & B & 34\\
TT Boo(2004 late) & 0.07796 & $-$6.2(0.9) & B & 34\\
RZ Leo      & 0.07853 &  5.9(1.0)  & WZ & 35\\
SU UMa      & 0.0788  & $-$10(3)     & -- & 36\\
HS Vir      & 0.08077 & $-$4(1)      & -- & 37\\
V877 Ara    & 0.08411 & $-$14.5(2.1) & -- & 38\\
EF Peg(1991)& 0.0871  & $-$2.2(1)    & B  & 39\\
BF Ara      & 0.08797 & $-$0.8(1.4)  & -- & 40\\
KK Tel      & 0.08801 & $-$37(4)     & -- & 7\\
V344 Lyr    & 0.09145 & $-$0.8(0.4)  & -- & 41\\
YZ Cnc      & 0.09204 & $-$7(2)      & -- & 42\\
V725 Aql    & 0.09909 & $\sim$0         & -- & 43\\
MN Dra   & 0.10768 & $-$170(2)    & -- & 44\\
TU Men      & 0.1262  & $-$9(2)      & -- & 45\\
XTEJ1118+480& 0.17053 & $-$0.6(0.1)  & A  & 46\\
\end{longtable}

\end{document}